# Is It Possible to Predict Strong Earthquakes?

Polyakov Y.S.[1,2], Ryabinin G.V.[3], Solovyeva A.B.[4], Timashev S.F.[5,6]

**Abbreviated Title**: Is It Possible to Predict Strong Earthquakes?

**Abstract** The possibility of earthquake prediction is one of the key open questions in modern geophysics. We propose an approach based on the analysis of common short-term candidate precursors (2 weeks to 3 months prior to strong earthquake) with the subsequent processing of brain activity signals generated in specific types of rats (kept in laboratory settings) who reportedly sense an impending earthquake few days prior to the event. We illustrate the identification of short-term precursors using the groundwater sodium-ion concentration data in the time frame from 2010 to 2014 (a major earthquake occurred on February 28, 2013), recorded at two different sites in the south-eastern part of the Kamchatka peninsula, Russia. The candidate precursors are observed as synchronized peaks in the nonstationarity factors, introduced within the flicker-noise spectroscopy framework for signal processing, for the high-frequency component of both time series. These peaks correspond to the local reorganizations of the underlying geophysical system that are believed to precede strong earthquakes. The rodent brain activity signals are selected as potential "immediate" (up to 2 weeks) deterministic precursors due to the recent scientific reports confirming that rodents sense imminent earthquakes and the population-genetic model of Kirshvink (2000) showing how a reliable genetic seismic escape response system may have developed over the period of several hundred million years in certain animals. The use of brain activity signals, such as electroencephalograms, in contrast to conventional abnormal animal behavior observations, enables one to apply the standard "input-sensor-response" approach to determine what input signals trigger specific seismic escape brain activity responses.

[1] New Jersey Institute of Technology, Newark, NJ, USA. E-mail: polyakov@njit.edu
[2] USPolyResearch, Ashland, PA, USA.
[3] Kamchatka Branch, Geophysical Survey of Russian Academy of Sciences, Petropavlovsk-Kamchatsky, Russia.
[4] Semenov Institute of Chemical Physics, Russian Academy of Sciences, Moscow Russia.
[5] Institute of Laser and Information Technologies, Russian Academy of Sciences, Troitsk, Moscow Region, Russia.
[6] National Research Nuclear University MEPhI, Moscow, Russia.

*1. Introduction*

Earthquake prediction within a time frame of several months to less than an hour before the catastrophic event, which is often referred in literature as "short-term" prediction, has been a subject of extensive research studies and controversial debates, both in academia and mass media, in the past two decades (Geller, 1997; Geller *et al.*, 1997; Wyss *et al.*, 1997; Uyeda *et al.,* 2009; Cicerone *et al.*, 2009). A recent striking example is when Italian seismologists were prosecuted in court and given a six-year jail sentence for providing "inaccurate, incomplete and contradictory" statements regarding the risk of a major earthquake near L'Aquila, with a series of foreshocks recorded few days prior to the deadly seismic event (Hall, 2011; "Shock and law", 2012). One of the key areas in earthquake prediction is the study of precursors, physical phenomena that reportedly precede at least some earthquakes. The precursory signals are usually grouped into electromagnetic, hydrological/hydrochemical, gasgeochemical, geodetic, and seismic (Geller, 1997; Hartmann and Levy, 2005; Uyeda *et al.*, 2009; Cicerone *et al.*, 2009; Ryabinin *et al.,* 2011; Bormann, 2011). Abnormal animal behavior prior to strong earthquakes, which was based in the past mostly on anecdotal and retrospective evidence, has recently been reported and studied in scholarly journals, and may now be considered as a potential precursor as well (Yokoi *et al.*, 2003; Li *et al.*, 2009; Grant and Halliday, 2010; Lu-Lu *et al.*, 2010; Grant *et al.* 2011).

Despite the large number of earthquake precursors reported in literature, most of which are summarized by Hartmann and Levy (2005) and Cicerone *et al.* (2009), an International Commission on Earthquake Forecasting for Civil Protection concluded on 2 October 2009, "the search for precursors that are diagnostic of an impending earthquake has not yet produced a successful short-term prediction scheme" (ICEFCP, 2009). The reports of the International Association of Seismology and Physics of the Earth's Interior contain similar findings (Wyss and Booth, 1997). The lack of confidence can be attributed to several reasons. First, some fundamental aspects of many non-seismic signals, for example, lithosphere-atmosphere-ionosphere coupling and propagation of ultra low frequency electromagnetic signals in the conductive earth, are unresolved, and many of the proposed physical models are questionable (Uyeda *et al.*, 2009). Second, the experimental data on precursory signals are often limited to few earthquakes and few measurement sites and they frequently contain gaps and different types of noise (Hartmann and Levy, 2005; Cicerone *et al.*, 2009, Uyeda *et al.* 2009). Third, different techniques of identifying the anomalies are used for different signals or even in different studies for the same signal. In some cases, the anomalous changes are determined by analyzing the signals themselves (Hartmann and Levy, 2005; Uyeda *et al.*, 2009; Cicerone *et al.*, 2009), while in other cases they are identified by studying the derived statistics or functions, such as Fisher information or scaling parameters (Telesca *et al.*, 2009a,b). Moreover, seasonal changes and instrumentation or other background noise often need to be filtered out prior to the identification of precursors.

To overcome some of these challenges, we proposed a phenomenological approach to searching for earthquake precursors based on the analysis of signals of different types in the same local geographic region (Ryabinin e*t al.*, 2011, 2012). We assume that a large earthquake may be preceded by a reconfiguration of a geophysical system on different time and space scales, which manifests itself in qualitative changes of various signals within relatively short time intervals. Our previous studies performed using flicker-noise spectroscopy (FNS), a phenomenological framework for extracting information from time series with stochastically varying components (Timashev and Polyakov, 2007, 2008; Timashev, 2007; Timashev *et al.*, 2010a), show that the peak values in FNS nonstationarity factors, which correspond to the time moments of major rearrangements (within relatively short time intervals) of a complex geophysical system (preceding a future strong earthquake), may be considered as "precursors" of the upcoming earthquakes (Descherevsky *et al.*, 2003; Telesca *et al.*, 2004; Vstovsky *et al.*, 2005; Hayakawa and Timashev, 2006; Ida *et al.*, 2007; Ryabinin e*t al.*, 2011). These studies demonstrate that the timing of the peaks in FNS nonstationarity factors with respect to strong earthquakes may dramatically vary depending on the nature of the signals under study. In this regard, it was suggested to introduce "short-term" (weeks to months prior to the earthquake) and "immediate" (minutes to weeks) precursors (Ryabinin *et al.*, 2011). The short-term precursors may be identified by examining the nonstationarity factor for daily time series of a certain parameter recorded in a seismically active zone; for instance, it can be water salinity in a borehole within that geographical area (Ryabinin *et al.*, 2011; 2012). The immediate precursors can be found by analyzing the nonstationarity factor for hourly, and then minute-level, sequences of another appropriately selected dynamic variable, such as electrotelluric or electrochemical potential, intensity of geoacoustic or electromagnetic signals in the ultra low frequency band. A combined analysis of these two kinds of precursors could be used to forecast a strong earthquake (Ryabinin *et al.*, 2011; 2012).

One can argue that such forecast would only be probabilistic rather then deterministic. The occurrence of the peak value in the nonstationarity factor reflects some major reconfiguration in the geophysical system, which may be treated only as a necessary but not a sufficient condition of a strong earthquake. For instance, the geophysical medium may simultaneously experience relaxation rearrangements, which would also be seen as peak values in the nonstationarity factor. This reasoning can be applied to interpret Figs. 2 and 3 in (Ryabinin *et al.*, 2011). The question is how to differentiate between the abrupt changes preceding a strong earthquake from these relaxation phenomena, and deterministically predict the earthquake. In other words, what could be used as a "sufficient" condition of the upcoming strong earthquake? Can such sufficient condition be formulated to enable earthquake prediction?

In this study, we attempt to answer these questions. First, we illustrate the procedure for identifying a necessary condition for a strong earthquake based on the analysis of hydrogeochemical time series for the Kamchatka peninsula in the time frame from 2010 to 2014 (a major earthquake was recorded on February 28, 2013), and then we discuss our hypothesis regarding the selection of an appropriate sufficient condition.

The paper is structured as follows. In Section 2, we provide the fundamentals of FNS and present the nonstationarity factor. Section 3 describes the experimental setup used to illustrate the selection of a necessary condition. Section 4 deals with the nonstationarity analysis of the hydrogeochemical data. Section 5 discusses the results and elaborates on the mechanism for selecting the sufficient condition. Section 6 presents the concluding remarks.

*2. FNS Nonstationarity Factor*

Here, we will only deal with the basic FNS relations needed to understand the nonstationarity factor. FNS is described in more detail elsewhere (Timashev and Polyakov, 2007, 2008; Timashev, 2007; Timashev *et al.*, 2010a, 2012; Polyakov *et al.*, 2012).

In FNS, all introduced parameters for signal *V*(*t*), where *t* is time, are related to the autocorrelation function

$$\psi(\tau) = \langle V(t)V(t+\tau)\rangle_T, \qquad (1)$$

where $\tau$ is the time lag parameter ($0 \leq \tau \leq T_M$) and $T_M$ is the upper bound for $\tau$ ($T_M \leq T/2$). This function characterizes the correlation in the values of dynamic variable *V* at higher, *t* + $\tau$, and lower, *t*, values of the argument. The angular brackets in relation (1) stand for the averaging over time interval [0,*T*]

$$\langle(...)\rangle_T = \frac{1}{T}\int_0^T (...)\,dt. \qquad (2)$$

The averaging over interval [0,*T*] implies that all the characteristics that can be extracted by analyzing functions $\psi(\tau)$ should be regarded as average values on this interval.

To extract the information contained in $\psi(\tau)$ ($\langle V(t) \rangle = 0$ is assumed), the following transforms, or "projections", of this function are analyzed: cosine transforms ("power spectrum" estimates) $S(f)$, where $f$ is the frequency,

$$S(f) = 2 \int_0^{T_M} \langle V(t)V(t+t_1) \rangle_{T-\tau} \cos(2\pi f t_1) \, dt_1 \qquad (3)$$

and its difference moments (Kolmogorov transient structure functions) of the second order $\Phi^{(2)}(\tau)$

$$\Phi^{(2)}(\tau) = \langle [V(t) - V(t+\tau)]^2 \rangle_T. \qquad (4)$$

Here, we use the quotes for power spectrum because according to the Wiener-Khinchin theorem the cosine (Fourier) transform of autocorrelation function is equal to the power spectral density only for wide-sense stationary signals at infinite integration limits.

The information contents of $S(f)$ and $\Phi^{(2)}(\tau)$ are generally different, and the parameters for both functions are needed to solve parameterization problems. By considering the intermittent character of signals under study, interpolation expressions for the stochastic components $S_s(f)$ and $\Phi_s^{(2)}(\tau)$ of $S(f)$ and $\Phi^{(2)}(\tau)$, respectively, were derived using the theory of generalized functions by Timashev (2006). It was shown that structural functions $\Phi_s^{(2)}(\tau)$ are formed only by jump-like (random-walk) irregularities corresponding to a dissipative process of anomalous diffusion, and functions $S_s(f)$, which characterize the "energy side" of the process, are formed by spike-like (inertial) and jump-like irregularities. It should be noted that $\tau$ in Eqs. (1)-(4) is considered as a macroscopic parameter exceeding the sampling period by at least one order of magnitude. This constraint is required to derive the expressions and separate out the contributions of dissipative jump-like and inertial (non-dissipative) spike-like components.

The analysis of experimental stochastic series often requires the original data to be smoothed. In this study, we will apply the "relaxation" procedure proposed for nonstationary signals by Timashev and Vstovskii (2003) based on the analogy with a finite-difference solution of the diffusion equation, which allows one to split the original signal into low-frequency $V_R(t)$ and high-frequency $V_F(t)$ components. The iterative procedure finding the new values of the signal at every "relaxation" step using its values for the previous step allows one to determine the low-frequency component $V_R$ (see expressions (A.7)-(A.9) in Appendix A). The high-frequency component $V_F$ is obtained by subtracting $V_R$ from the original signal. This algorithm progressively reduces the local gradients of the "concentration" variable, causing the points in every triplet to come closer to each other. The smoothing procedure is described in more detail in Appendix A.

To analyze the effects of nonstationarity in real processes, we study the dynamics of changes in $\Phi^{(2)}(\tau)$ for consecutive "window" intervals $[t_k, t_k+T]$, where $k = 0, 1, 2, 3, \ldots$ and $t_k = k\Delta T$, that are shifted within the total time interval $T_{tot}$ of experimental time series ($t_k+T < T_{tot}$). The averaging interval $T$ and difference $\Delta T$ are chosen based on the physical understanding of the problem in view of the suggested characteristic time of the process, which is the most important parameter of the system evolution. The FNS nonstationarity factor is defined as:

$$C_J(t_k) = 2 \cdot \frac{Q_k^J - P_k^J}{Q_k^J + P_k^J} \bigg/ \frac{\Delta T}{T}, \tag{5}$$

$$Q_k^J = \frac{1}{\alpha T^2} \int_0^{\alpha T} \int_{t_k}^{t_k+T} [V_J(t) - V_J(t+\tau)]^2 \, dt \, d\tau, \tag{6}$$

$$P_k^J = \frac{1}{\alpha T^2} \int_0^{\alpha T} \int_{t_k}^{t_k+T-\Delta T} [V_J(t) - V_J(t+\tau)]^2 \, dt \, d\tau. \tag{7}$$

Here, $J$ indicates which function $V_J(t)$ ($J = R, F$ or $G$) is used to evaluate $\Phi_J^{(2)}(\tau)$, and the subscripts $R, F,$ and $G$ refer to the low-frequency component, high-frequency component, and unfiltered signal, respectively. Expressions (5)-(7) are given in discrete form in Appendix B. Note that functions $\Phi_J^{(2)}(\tau)$ can be reliably evaluated only on the $\tau$ interval of $[0, \alpha T]$, which is less than half the averaging interval $T$; i.e., $\alpha < 0.5$.

The phenomenon of "precursor" occurrence is assumed to be related to abrupt changes in functions $\varPhi^{(2)}(\tau)$ when the upper bound of the interval $[t_k, t_k+T]$ approaches the time moment $t_c$ of a catastrophic event accompanied by total system reconfiguration on all space scales. Graphically, this corresponds to peaks in the plots of nonstationarity factor.

3. Data

The data were recorded in the south-eastern part of the Kamchatka peninsula located at the Russian Far East. The eastern part of the peninsula is one of the most seismically active regions in the world. The area of highest seismicity localized in the depth range between 0 and 40 km represents a narrow stripe with a length of approximately 200 km along the east coast of Kamchatka, which is bounded by a deep-sea trench on the east (Fedotov *et al.*, 1985).

Specialized measurements of groundwater characteristics were started in 1977 to find and study possible hydrogeochemical precursors of Kamchatka earthquakes. Currently, the observation network includes four stations in the vicinity of Petropavlovsk-Kamchatsky: Pinachevo, Moroznaya, Khlebozavod, and Verkhnyaya Paratunka. The Pinachevo station includes five water reservoirs: four warm springs and one borehole GK-1 with a depth of 1,261 m. The Moroznaya station has a single borehole No. 1 with a depth of 600 m. The Khlebozavod station also includes a single borehole G-1 with a depth of 2,540 m, which is located in Petropavlovsk-Kamchatsky. The Verkhnyaya Paratunka station comprises four boreholes (GK-5, GK-44, GK-15, and GK-17) with depths in the range from 650 to 1208 m.

The system of hydrogeochemical observations includes the measurement of atmospheric pressure and air temperature, measurement of water discharge and temperature of boreholes and springs, and collection of water and gas samples for their further laboratory analyses. For water samples, the following parameters are determined: pH; ion concentrations of chlorine ($Cl^-$), bicarbonate ($HCO_3^-$), sulfate ($SO_4^{2-}$), sodium ($Na^+$), potassium ($K^+$), calcium ($Ca^{2+}$), and magnesium ($Mg^{2+}$); concentrations of boric ($H_3BO_3$) and silicone ($H_4SiO_4$) acids. For the samples of gases dissolved in water, the following concentrations are determined: methane ($CH_4$), nitrogen ($N_2$), oxygen ($O_2$), carbon dioxide ($CO_2$), helium (He), hydrogen ($H_2$), hydrocarbon gases: ethane ($C_2H_6$), ethylene ($C_2H_4$), propane ($C_3H_8$), propylene ($C_3H_6$), butane ($C_4H_{10}n$), and isobutane ($C_4H_{10}i$). The data are recorded at nonuniform sampling intervals with one dominant sampling frequency. For the Pinachevo, Moroznaya, and Khlebozavod stations, this average sampling frequency is one measurement per 3 days; for the Verkhnyaya Paratunka station, one measurement per 6 days. Multiple studies of the hydrogeochemical data and corresponding seismic activity for the Kamchatka peninsula reported anomalous changes in the chemical and/or gas composition of groundwater prior to several large earthquakes (Kopylova *et al.*, 1994; Bella *et al.*, 1998; Biagi *et al.* 2000, 2004, 2006; Khatkevich and Ryabinin, 2006).

In this study, we analyze the time series at GK-1 (Pinachevo station) and G-1 (Khlebozavod station). Our goal is to examine the buildup for the major earthquake of February 28, 2013 and subsequent earthquake swarm in May of 2013, which are illustrated in Fig. 1.

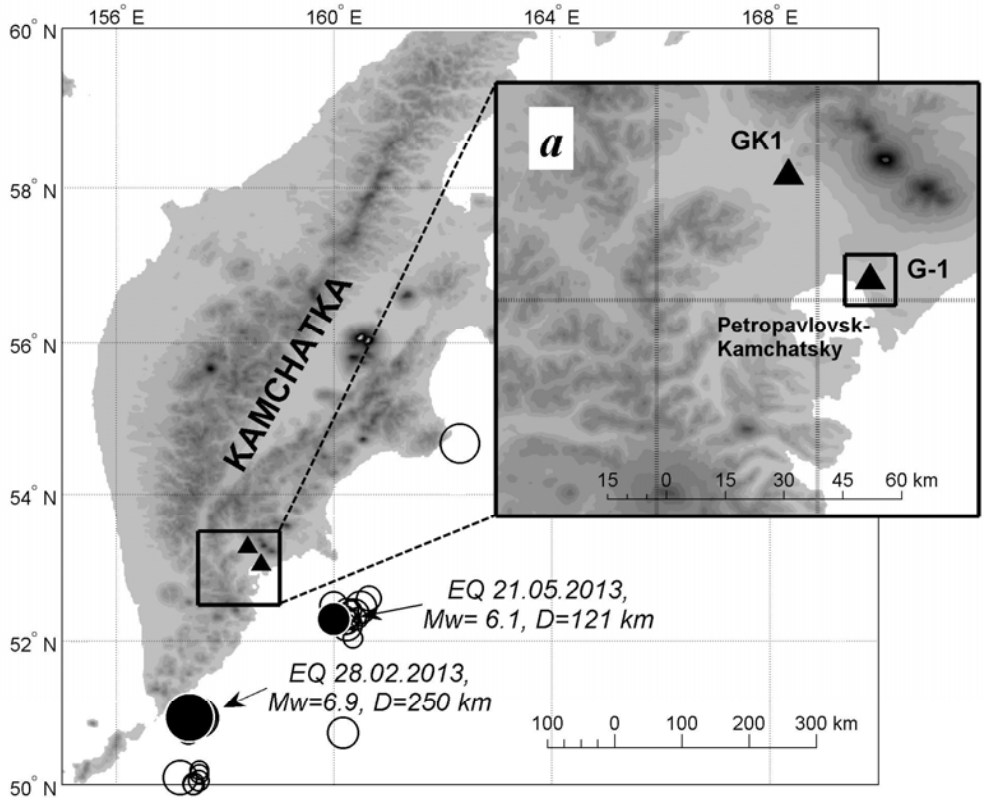

Fig. 1. Schematic of the measurement area, observation points, and epicenters of strongest earthquakes in 2013 ($M \geq 5.0$, $H \leq 50$ km, $D \leq 350$ km), where $M$ – earthquake magnitude, $H$ – depth, $D$ – distance from the epicenter to Petropavlovsk-Kamchatsky. Solid black circles denote the major earthquake of February 28, 2013 ($M_w$=6.9, $D$=250 km) and strongest earthquake ($M_w$=6.1, $D$=121 km) in a series of seismic events in May of 2013. Frame *a* shows an enlarged view of the positions of boreholes G-1 and GK-1. The earthquakes were selected using the catalog of U.S. Geological Survey: http://earthquake.usgs.gov/earthquakes/search/.

*4. Results*

To illustrate the procedure for identifying a necessary condition of the impending strong earthquake, we examined the signals in the time interval from January 1, 2010 to January, 2014. This period is characterized by one major earthquake with a magnitude above 6.5, which took place on February 28, 2013. According to our hypothesis that any large earthquake is preceded by a major reconfiguration of the geophysical system in the affected region (Ryabinin *et al.*, 2011), we expected to observe strong qualitative changes in at least some of the hydrogeochemical parameters occurring simultaneously at multiple measurement stations in the Kamchatka region. Our preliminary analysis showed that the nonstationarity factor for the high-frequency component of sodium-ion concentration shows a striking peak at multiple stations 52-54 days prior to the strongest earthquake (Figs. 2 and 3). As a result, we selected this hydrogeochemical parameter for further analysis.

Figures 2 and 3 display the variation of sodium-ion concentration in boreholes GK-1 and G-1, respectively, along with the nonstationarity factor $C_F$ evaluated at three different values of the averaging interval $T$ and seismic activity data during this period. The average sampling frequency of the sodium-ion concentration time series is $(3 \text{ day})^{-1}$. The high-frequency component was obtained by applying 10 iterations of the smoothing procedure (A.7)-(A.9) at $\omega=0.25$, which corresponds to the effective high-pass cutoff frequency of approximately $(18 \text{ day})^{-1}$ (where the value of power spectrum density for the component reaches its maximum).

It can be seen in Fig. 2 that the highest peak in the nonstationarity factor $C_F$ at $T = 75$ days for the sodium-ion concentration in borehole GK-1 occurs 52 days prior to the earthquake of February 28, 2013. As we increase the averaging interval $T$, the relative magnitude of the peak gets less pronounced due to the contribution of other local reorganizations in the high-frequency range during the current averaging interval $[t_k, t_k+T]$. This implies that the averaging interval of 75 days is most adequate for identifying single local reorganizations in the high-frequency component of the sodium-ion concentration data for GK-1, at least for the earthquake under study. It should also be noted that there is a relatively high peak in the nonstationarity factor approximately nine months prior to the major earthquake. This peak is not followed by any major seismic event, which suggests that not every local reorganization leads to a strong earthquake.

The nonstationarity factor $C_F$ in Fig. 3 has the highest peak 54 days prior to the strongest earthquake at all three values of $T$. Considering that the average sampling interval for both time series is 3 days, we can conclude that the nonstationarity factors for both boreholes point to the same local reorganization, which happens to be the most significant one in the time period under study. It is reasonable to relate this synchronized local reorganization with a buildup for some major event, such as the earthquake of February 28. It is noteworthy that there is another noticeable peak in the nonstationarity factor $C_F$, which occurs approximately at the same time (nine months prior to the major earthquake) as in the data for GK-1.

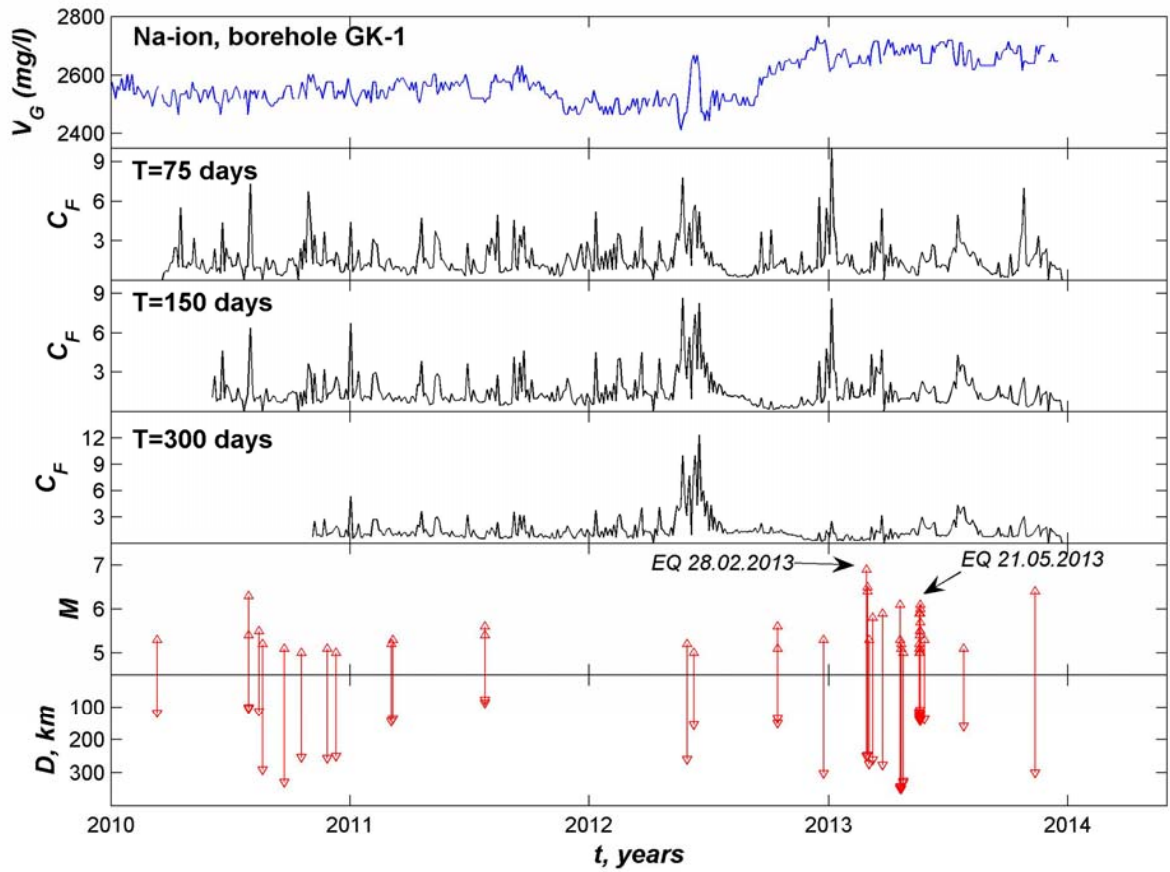

Fig. 2. Comparison of nonstationarity factor $C_F$ for the high-frequency component of sodium-ion concentration in the water of borehole GK-1 (Pinachevo station) with seismic activity ($M \geq 5.0$): $V_G$, unfiltered time series of sodium-ion concentration; $M$, earthquake magnitude; $D$, distance of the epicenter from Petropavlovsk-Kamchatsky.

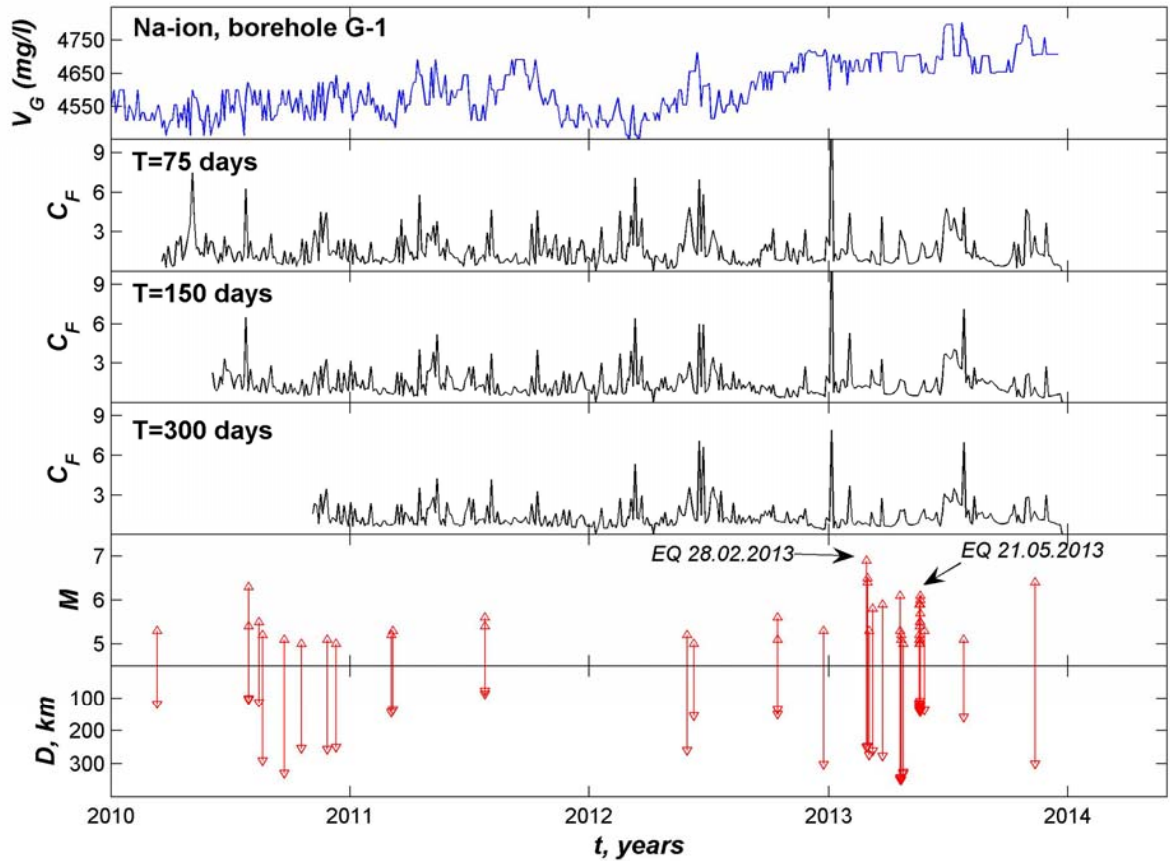

Fig. 3. Comparison of nonstationarity factor $C_F$ for the high-frequency component of sodium-ion concentration in the water of borehole G-1 (Khlebozavod station) with seismic activity ($M \geq 5.0$): $V_G$, unfiltered time series of sodium-ion concentration; $M$, earthquake magnitude; $D$, distance of the epicenter from Petropavlovsk-Kamchatsky.

## 5. Discussion

The largest peaks in FNS nonstationarity factors for the high-frequency component of groundwater sodium-ion concentration simultaneously occurring at two stations 32 km apart from each other suggest that a relatively large-scale reorganization may have preceded the earthquake of February 28, 2013. The peaks were observed 52-54 days prior to the major seismic event, which is agreement with our previous results for the variation of groundwater chlorine-ion concentration where the peaks were observed in the range from 50 to 70 days at GK-1 and GK-44 boreholes prior to the major earthquake of October 8, 2001 (Ryabinin *at al.*, 2011). Large peaks in the nonstationarity factor before strong earthquakes were also observed for other types of data, including geoelectrical signals (Telesca *et al.*, 2004), electrochemical potentials (Descherevsky *et al.*, 2003), and ultra-low-frequency electromagnetic transmissions (Hayakawa and Timashev, 2006). This implies that local reorganizations in the geophysical system of a particular seismic region may be related to a buildup for impending earthquakes.

At the same time, such local reorganizations (peaks in the nonstationarity factor) do not always lead to a strong earthquake, as illustrated by Figs. 2 and 3 in (Ryabinin *et al.*, 2011) and the other nonstationarity factor peaks observed in Figs. 2 and 3 of this study. These nonstationarity factor peaks may also be related to certain rearrangements when the geophysical system responds to abrupt changes in external conditions. It is also possible that a series of local reorganizations has to take place prior to triggering a strong earthquake, and only the last local reorganization could be considered in this case as a real precursor. The question is how to differentiate between the abrupt changes preceding a strong earthquake from these non-precursory local rearrangements. Taking into account the low number of major earthquakes for which the accompanying non-seismic time series are available and the absence of reliable physical models explaining the correlation between the variability in the datasets and seismic events, it can be concluded that such distinction cannot be made at this time and the peaks in nonstationarity factors of certain signals cannot be regarded as deterministic precursors. The occurrence of the peak can only imply that a major earthquake is possible within a certain interval of time after the peak (in the case of hydrogeochemical signals with a sampling frequency of $(3 \text{ day})^{-1}$, the strong earthquake may be expected to happen between 50 and 70 days after the peak). In other words, a high peak in the nonstationarity factor can only be regarded as a necessary condition for a strong seismic event. Is it possible to complement this necessary condition by a sufficient condition to enable true earthquake prediction?

We believe that a truly reliable system for predicting strong earthquakes may have developed over the period of several hundred million years in certain animals, which is supported by several recent scientific studies reporting abnormal behavior in mice, rats, and toads in the range from 1 to 6 days before strong earthquakes (Yokoi *et al.*, 2003; Li *et al.*, 2009; Grant and Halliday, 2010; Lu-Lu *et al.*, 2010; Grant *et al.* 2011). Kirshvink (2000) argues that the evolutionary mechanism of exaptation (the ability to adapt or link a genetic pattern that evolved for one function for another role) may have produced a seismic-response system through the process of random mutation and natural selection. The facts that first burrowing animals appeared at least 540 million years ago and the plate tectonics on Earth has operated for at least two billion years suggest that from the distant ancestors of all mammals (synapsids) up to modern age mammals the evolution dealt with a very large sample of strong earthquakes associated with increased mortality of the animals without appropriate seismic response gene(s). Using the concept of episodic selection and Monte-Carlo simulations, Kirshvink (2000) demonstrates that such hypothetical signal response gene could easily reach fixation within a population only after 1000 generations, with every seismic event occurring once in 50 generations and resulting in 10% mortality of the individuals without the gene. In contrast, the sample of strong earthquakes and accompanying seismic and non-seismic data that are available to the scientific community is by many orders of magnitude smaller, which makes many seismologists believe that earthquake prediction is impossible (Geller, 1997; Geller *et al.*, 1997). Another possible explanation discussed in literature is that the anomalous animal behavior observed prior to strong earthquakes is a direct response of animals to certain adverse stimuli caused by the changes in the environment shortly before the seismic event (Grant *et al.* 2011; Freund and Stolc 2013). These stimuli may be brought about by changes in the concentrations of $H_2O_2$ in water, CO in air, or positive ions in air; shifts in electromagnetic fields; or certain variations in other physical/chemical characteristics (Freund and Stolc 2013).

Up to the present, abnormal animal behavior has not been considered as a precursor candidate by many seismologists and was not submitted for evaluation to the IASPEI Sub-Commission on Earthquake

Prediction because it cannot be permanently monitored in a controlled way and unambiguously assessed (Bormann, 2011). It is generally believed that similar anomalous animal behavior may have different causes and cannot be reproduced using the standard "input-sensor-response" approach. This reasoning is based on the fact that many past studies relied on subjective behavioral observations, such as anomalous locomotive activities in the circadian rhythms of mice (Yokoi *et al.*, 2003; Li *et al.*, 2009) or the quiescence or lack of spawning in toads (Grant and Halliday, 2010). However, if we rely on more objective biomedical signals and study the responses to certain stimuli believed to be related to earthquake buildup, we can apply the standard input-sensor-response method and identify precursors by analyzing the signals rather than the general behavior of the animals under study. One potential approach to obtaining more objective signals may be based on the computer-aided measurements of animal behavior with specialized video cameras, automated logging machines, motion-triggered cameras, or the use of other automated technologies for recording anomalous animal responses. However, such automatically recorded behavioral responses would only capture the increase or decrease in certain parameters of locomotive activity, thus ignoring the complexity of the biological system of the animal and greatly reducing the number of degrees of freedom. It is rather likely that two different stimuli (precursory and non-precursory) may trigger the same change in the selected parameter of locomotive activity. Therefore, it is necessary to look beyond behavioral "signals" and select candidate signals that better represent the complexity of the biological system.

In our view, the brain activity signals in animals can be considered as the main biomedical data because the brain, operating as an integral system, processes any external signals, including those related to imminent danger, and generates specific responses that can be identified from the analysis of cortical brain activity. In this regard, electroencephalogram (EEG) measurements of relatively large rodents, such as California kangaroo rats, obtained with two-channel implantable EEG sensor transmitters (manufactured by BIOPAC Systems Inc.) or 4-channel tethered systems with rat headmounts (manufactured by Pinnacle Technology Inc.) appear to be the most practical option for seismological studies. Small laboratories with large enough rat samples in each could be set up at multiple sites of a specific seismically active region, and the online analysis of the EEG data could be performed in a central signal processing facility. The EEG measurements in rats have to be performed in accordance with established animal ethical guidelines. A potential challenge of this method is the placement of electrodes (Nobler *et al.* 1993), which may require initial calibration to capture earthquake-relevant responses in the rodents.

It should be noted that the FNS framework looks promising for analyzing these EEG data because it recently demonstrated its effectiveness in the analysis of human EEG and magnetoencephalogram (MEG) signals. The FNS parameterization procedure and cross-correlation function were used to classify the susceptibility of 84 children/adolescents (11 to 14 years of age) to schizophrenia based on the analysis of EEG signals (Timashev *et al.*, 2012). As part of this study, three new quantitative diagnostic markers were suggested. In another study, the FNS parameterization procedure and cross-correlation function were effective in studying neuromagnetic responses (MEG signals) from a group of healthy human subjects (9 volunteers) and a photosensitive epilepsy (PSE) patient while they were viewing equiluminant flickering stimuli of different color combinations (Timashev *et al.*, 2009, 2010b). The purpose of that study was to develop diagnostic biomarkers for PSE. These results along with the findings reported by Ryabinin *et al.* (2011, 2012) imply that all three major FNS tools (parameterization procedure, cross-correlation function, nonstationarity factor) may be applied in the analysis of rat EEG signals.

For an earthquake prediction method to be successful, it should specify: (1) the time window, (2) the spatial window; (3) the magnitude window (Geller, 1997). According to Kirshvink (2000), Tributsch (1984), and Grant and Halliday (2010), animals can generally sense an impending strong earthquake at most 1-2 weeks prior to the event, with highest abnormal activity occurring 1-2 days before the earthquake (here, we ignore the ability of animals to sense the *P* waves traveling immediately before the event because the time interval of few seconds is of no practical interest). This provides a relatively narrow time window for the imminent seismic event. The spatial window can be determined by processing the EEG data from multiple sites; we expect there should be some synchronization between the EEG patterns of rats in the geographical zone where the effect of the strong earthquake would be detrimental. The magnitude window in this case could not be given directly in terms of magnitude and depth, but the presence of specific EEG patterns related to the seismic escape behavior implies that the mortality of the rodents under study could be affected. In other words, only relatively strong earthquakes (with magnitude $M > 5$) would be predicted by this method. The above suggests that if such system can be built and appropriate EEG patterns identified, it may be adequate for the deterministic prediction of strong earthquakes.

It is evident that the nature of external signals warning the animals about the upcoming seismic event is yet to be determined. Kirshvink (2000) suggested that there are four kinds of "precursory" signals that may be sensed by animals: (1) ground tilting, (2) humidity changes, (3) electrical currents, and (4) magnetic fields variations. There are no experimental data supporting that ground tilting could be the main input signal. Humidity changes can only be sensed in arid environments, which does not explain why mice, rats, and toads in non-desert areas demonstrated abnormal behavior prior to a number of strong earthquakes. Electrical currents could probably be sensed only by aquatic animals, whose electrical sensitivity is much higher than in terrestrial animals. The only relatively universal mechanism observed in many vertebrate animals is a highly sensitive magnetoreception system that is capable to sense ultra-low frequency magnetic activity known to precede certain earthquakes because this system probably evolved prior to the divergence of major subgroups of vertebrates (Kirshvink, 2000). This system was most likely developed for navigational and/or circadian purposes, and later may have been adapted through exaptation for the seismic escape response role (Kirshvink, 2000). If the precursory signals sensed by the animals are of biomagnetic nature, then the proposed EEG experiments should consider the negative effect of radio frequency noise on the magnetoreception system of the rodents (Kirshvink, 2014). Grant *et al.* (2011) and Freund and Stolc (2013) suggested that the activation of mobile electronic charge carriers caused by the buildup of tectonic stress may generate electromagnetic signals in the ultra low frequency (ULF) and extremely low frequency (ELF) ranges. These signals, along with resulting positively charged ions in air and elevated levels of hydrogen peroxide in water, may cause unusual reactions in certain terrestrial and aquatic animals. There may be another kind of precursory signals that could be sensed by certain animals: low-frequency geoacoustic noise. Kirshvink (2000) discussed in detail the ability of many animals to sense acoustic *P* (primary) waves that are detected few seconds before the destructive *S* (secondary) waves arrive. At the same time, some studies show that geoacoustic precursors, such as peaks in the nonstationary factors for geoacoustic emissions in the frequency range from 25 to 1,400 Hz recorded in a borehole of the Khlebozavod station (Ryabinin *et al.*, 2011), are observed several days prior to certain earthquakes. If animals can sense the *P* waves, maybe they could also detect the anomalous changes in such low-frequency geoacoustic noise.

In contrast to behavioral observations, the analysis of rat EEG signals also enables one to perform the laboratory experiments where brain activity responses to specific types of input are studied and categorized in an objective way. First, controlled shake-table experiments could be conducted using laboratory populations of rats from seismically active zones, for example, California kangaroo rats. These experiments would establish a baseline of EEG response patterns for comparison with reactions to other stimuli. Then a series of field-based experiments could be performed on the same species in which candidate precursory geophysical and geochemical signals are applied artificially to determine if any of the input signals would produce similar EEG patterns. The exaptation model developed by Kirshvink (2000) would suggest some similarity in the evoked EEG response between shaking and other stimuli reported to precede certain strong earthquakes. This input-sensor-output analysis would help understand what earthquake precursory signals are sensed by the rats.

*5. Concluding remarks*

The above discussion suggests that an earthquake prediction system might be built using a combined approach comprising (1) the identification of short-term candidate precursors (2 weeks to several months prior to major seismic event) by locating the peaks in FNS nonstationarity factors of relatively low-frequency (hourly, daily) signals manifesting local reorganizations in the underlying geophysical system prior to strong earthquakes, and (2) the subsequent processing of EEG signals of the rats living in seismic areas, for instance, California kangaroo rats, collected from their lab populations at multiple sites within the same seismically active region. We show that groundwater sodium-ion concentration with a sampling frequency of $(3 \text{ day})^{-1}$ can be chosen as a short-term candidate precursor for the earthquakes similar to the one of February 28, 2013 in the Kamchatka peninsula region. It should be noted that not every candidate precursor is followed by a strong earthquake and that the monitoring of rat EEG signals does not need to be performed continuously. The occurrence of the short-term candidate precursor should trigger a detailed EEG analysis within the 3 months following the peak in nonstationarity factor (for the case studied in this paper). If no EEG patterns corresponding to a specific animal seismic response are detected within that time frame, the EEG signal monitoring can be stopped until the next peak in nonstationarity factor for the short-term candidate precursory signal is identified.

Laboratory experiments where EEG responses to specific types of input are studied and compared to the responses in controlled shake-table experiments should be conducted to identify the signals detected by the sensory system of the rats prior to strong seismic events. The brain activity responses and input sensory signals may vary depending on the characteristics of impending strong earthquakes. For instance, shallow-focus, mid-focus, and deep-focus earthquakes may each be characterized by different precursory signals (for example, electromagnetic emissions in different frequency ranges). The ultimate goal of laboratory experiments would be to learn from the seismic escape response system of rats and build artificial signal receiving systems with sensors and centralized signal processing.

APPENDIX A

Consider the one-dimensional diffusion equation for $V_R$:

$$\frac{\partial V_R}{\partial \tau} = \chi \frac{\partial^2 V_R}{\partial t^2} \tag{A.8}$$

with symmetry boundary conditions

$$\frac{\partial V_R}{\partial t} = 0 \text{ at } t = 0, \tag{A.9}$$

$$\frac{\partial V_R}{\partial t} = 0 \text{ at } t = T \tag{A.10}$$

and initial condition

$$V_R(0) = V(t), \tag{A.11}$$

where $\chi$ is a constant diffusion coefficient.

Writing a forward difference for the local term and second-order central difference for the diffusion term, Equation (A.8) gets transformed to

$$\frac{V_k^{i+1} - V_k^i}{\Delta \tau} = \chi \frac{V_{k+1}^i - 2V_k^i + V_{k-1}^i}{(\Delta t)^2}, \tag{A.12}$$

where $i$ is the "time" index and $k$ is the "spatial" index. Here, the subscript "$R$" is dropped for simplicity.

After introducing $\omega = \frac{\chi \Delta \tau}{(\Delta t)^2}$, Equation (A.12) can be further transformed to the following explicit finite difference expression:

$$V_k^{i+1} = \omega V_{k+1}^i + (1 - 2\omega) V_k^i + \omega V_{k-1}^i. \tag{A.13}$$

Analogously, the finite difference formulation for the complete problem (A.8)-(A.11) can be written as

$$V_1^{i+1} = (1 - 2\omega) V_1^i + 2\omega V_2^i, \tag{A.14}$$

$$V_k^{i+1} = \omega V_{k+1}^i + (1 - 2\omega) V_k^i + \omega V_{k-1}^i \quad \text{for } 1 < k < N, \tag{A.15}$$

$$V_N^{i+1} = (1-2\omega)V_N^i + 2\omega V_{N-1}^i. \tag{A.16}$$

Here, $i$ is the current iteration number and $N$ is the length of the time series. The smoothing procedure (A.14)-(A.16) is unconditionally stable for $\omega < 1/2$, which is the maximum allowed value for $\omega$ in the smoothing algorithm. There are two input parameters: the number of iterations $i_{max}$ (largest value for $i$) and the value for $\omega$. The smoothing procedure evaluates expressions (A.14)-(A.16) at each iteration.

APPENDIX B

The FNS nonstationarity factor in discrete for is written as ($b = \lfloor \Delta T / \Delta t \rfloor$, $N_1 = \lfloor \alpha N \rfloor$):

$$C_J(t_k) = 2 \cdot \frac{Q_k^J - P_k^J}{Q_k^J + P_k^J} \bigg/ \frac{\Delta T}{T}, \tag{B.1}$$

$$Q_k^J = \frac{1}{N_1} \sum_{n_\tau=1}^{N_1} \frac{1}{N - n_\tau} \sum_{m=1+kb}^{N-n_\tau+kb} \left[V_J(m) - V_J(m+n_\tau)\right]^p, \tag{B.2}$$

$$P_k^J = \frac{1}{N_1} \sum_{n_\tau=1}^{N_1} \frac{1}{N - n_\tau} \sum_{m=1+kb}^{N-n_\tau+(k-1)b} \left[V_J(m) - V_J(m+n_\tau)\right]^p. \tag{B.3}$$

REFERENCES

Bella, F., Biagi, P.F., Caputo, M., Cozzi, E., Monica, G.D., Ermini, A., Gordeev, E.I., Khatkevich, Y.M., Martinelli, G., Plastino, W., Scandone, R., Sgrigna, V., and Zilpimiani, D. (1998), *Hydrogeochemical anomalies in Kamchatka (Russia)*, Phys. Chem. Earth *23*, 921–925, doi: 10.1016/S0079-1946(98)00120-7.

Biagi P.F., Castellana, L., Piccolo, R., Minafra, A., Maggipinto, G., Ermini, A., Capozzi, V., Perna, G., Khatkevich, Y.M., and Gordeev, E.I. (2004), *Disturbances in groundwater chemical parameters* *related to seismic and volcanic activity in Kamchatka (Russia)*, Nat. Hazards Earth Syst. Sci. *4*, 535–539, doi: 10.5194/nhess-4-535-2004.

Biagi, P.F., Castellana, L., Minafra, A., Maggipinto, G., Maggipinto, T., Ermini, A., Molchanov, O., Khatkevich, Y.M., and Gordeev, E.I. (2006), *Groundwater chemical anomalies connected with the Kamchatka earthquake (M=7.1) on March 1992*, Nat. Hazards Earth Syst. Sci. *6*, 853–859, doi: 10.5194/nhess-6-853-2006.

Biagi, P.F., Ermini, A., Cozzi, E., Khatkevich, Y.M., and Gordeev, E.I. (2000), *Hydrogeochemical precursors in Kamchatka (Russia) related to the strongest earthquakes in 1988–1997*, Nat. Hazards *21*, 263–276, doi: 10.1023/A:1008178104003.

Bormann, P. (2011), *From earthquake prediction research to time-variable seismic hazard assessment applications*, Pure Appl. Geophys. *168* (2011), 329–366, doi 10.1007/s00024-010-0114-0.

Cicerone, R.D., Ebel, J.E., and Britton, J. (2009), *A systematic compilation of earthquake precursors*, Tectonophysics *476*, 371–396, doi: 10.1016/j.tecto.2009.06.008.

Descherevsky, A.V., Lukk, A.A., Sidorin, A.Ya., Vstovsky, G.V., Timashev, S.F. (2003), *Flicker-noise spectroscopy in earthquake prediction research*, Nat. Hazards Earth Syst. Sci.. *3* (3/4), 159-164, doi: 10.1016/j.pce.2003.09.017.

Fedotov, S.A., Gusev, A.A., Shumilina, L.S., and Chernyshova, V.G. (1985), *The seismofocal zone of Kamchatka* (in Russian), Vulkanologiya i Seismologiya (4), 91–107.

Freund, F., and Stolc, V. (2013), *Nature of pre-earthquake phenomena and their effects on living organisms*, Animals 3(2), 513-531; doi: 10.3390/ani3020513.

Geller, R.J. (1997), *Earthquake prediction: A critical review*, Geophys. J. Int. *131*, 425–450, doi: 10.1111/j.1365-246X.1997.tb06588.x.

Geller, R.J., Jackson, D.D., Kagan, Y.Y., and Mulargia, F. (1997), *Earthquakes cannot be predicted*, Science *275*, 1616–1617, doi: 10.1126/science.275.5306.1616.

Grant, R. A., and Halliday, T. (2010), *Predicting the unpredictable; evidence of pre-seismic anticipatory behaviour in the common toad*, J. Zool. *281*, 263–271, doi: 10.1111/j.1469-7998.2010.00700.x.

Grant, R.A., Halliday, T., Balderer, W.P., Leuenberger, F., Newcomer, M., Cyr, G., Freund, F.T. (2011), *Ground water chemistry changes before major earthquakes and possible effects on animals*, Int. J. Environ. Res. Public Health *8*, 1936-1956.

Hall, S.S. (2011), *Scientists on trial: At fault?*, Nature *477*, 264-269, doi:10.1038/477264a.

Hartmann, J. and Levy, J.K. (2005), *Hydrogeological and gasgeochemical earthquake precursors - A review for application*, Nat. Hazards *34*, 279–304, doi: 10.1007/s11069-004-2072-2.

Hayakawa, M. and Timashev, S.F. (2006), *An attempt to find precursors in the ULF geomagnetic data by means of Flicker Noise Spectroscopy*, Nonlin. Processes Geophys *13*, 255-263, doi:10.5194/npg-13-255-2006.

ICEFCP (2009), *Operational Earthquake Forecasting: State of Knowledge and Guidelines for Utilization*, http://www.iaspei.org/downloads/Ex_Sum_v5_THJ9_A4format.pdf, access: 4 February 2014.

Ida, Y., Hayakawa, M., and Timashev S. (2007), *Application of different signal analysis methods to the ULF data for the 1993 Guam earthquake*, Nat. Hazards Earth Syst. Sci. *7*, 479-487, doi:10.5194/nhess-7-479-2007.

Khatkevich, Y. and Ryabinin G. (2006), *Geochemical an ground-water studies in Kamchatka in the search for earthquakes precursors* (in Russian), Vulkanologiya i Seysmologiya (4), 34–42.

Kirschvink, J.L. (2000), *Earthquake prediction by animals: evolution and sensory perception*, Bull. Seismol. Soc. America *90*, 312-323, doi: 10.1785/0119980114.

Kirschvink, J.L. (2014), *Sensory biology: Radio waves zap the biomagnetic compass*, Nature *509*, 296-297, doi: 10.1038/nature13334.

Kopylova, G., Sugrobov, V., and Khatkevich, Y. (1994), *Variations in the regime of springs and hydrogeological boreholes in the Petropavlovsk polygon (Kamchatka) related to earthquakes* (in Russian), Vulkanologiya i Seysmologiya (2), 53–70.

Li, Y., Liu, Y., Jiang, Z., Guan, J., Yi, G., Cheng, S., Yang, B., Fu, T., and Wang, Z. (2009), *Behavioral change related to Wenchuan devastating earthquake in mice*, Bioelectromagnetics *30*, 613-620, doi: 10.1002/bem.20520.

Lu-Lu, Ch., Xiang, H., Juan, Zh., Hao-Hao, Zh., Wen, K., Wei-Hong, Y., Tian-Shu, Z., Jiao-Yue, Zh., and Ling, Y. (2010), *Increases in energy intake, insulin resistance and stress in rats before Wenchuan earthquake far from the epicentre*, Exp. Biol. Med. *235*, 1216-1223, doi: 10.1258/ebm.2010.010042.

Nobler, M. S., Sackeim, H.A., Solomou, M., Luber, B., Devanand, D.P, and Prudic, J. (1993), *EEG manifestations during ECT: effects of electrode placement and stimulus intensity*, Biol. Psych. 34, 321-330, doi: 10.1016/0006-3223(93)90089-V.

Polyakov, Yu. S., Neilsen, J., and Timashev, S. F. (2012), *Stochastic variability in x-ray emission from the black hole binary GRS 1915+105*, Astron. J. *143*, 148, doi:10.1088/0004-6256/143/6/148.

Ryabinin, G., Gavrilov, V.A., Polyakov, Yu.S., and Timashev, S.F. (2012), *Cross-correlation earthquake precursors in the hydrogeochemical and geoacoustic signals for the Kamchatka peninsula*, Acta Geophys. *60*, 874-893.

Ryabinin, G., Polyakov, Yu.S., Gavrilov, V.A., and Timashev, S.F. (2011), *Identification of earthquake precursors in the hydrogeochemical and geoacoustic data for the Kamchatka peninsula by flicker-noise spectroscopy*, Nat. Hazards Earth Syst. Sci. 11, 541-548, doi: 10.5194/nhess-11-541-2011.

*Shock and law* (2012), Editorial, Nature *490*, 446, doi:10.1038/490446b.

Telesca, L., Lapenna, V., Timashev, S., Vstovsky, G., Martinelli, G. (2004), *Flicker-Noise spectroscopy as a new approach to investigate the time dynamics of geoelectric signals measures in seismic areas*, Phys. Chem. Earth 29, 389-395.

Telesca, L., Lovallo, M., Ramirez-Rojas, A., and Angulo-Brown, F. (2009a), *A nonlinear strategy to reveal seismic precursory signatures in earthquake-related self-potential signals*, Physica A *388*, 2036–2040, doi: 10.1016/j.physa.2009.01.035.

Telesca, L., Lovallo, M., Ramirez-Rojas, A., and Angulo-Brown, F. (2009b), *Scaling instability in self-potential earthquake-related signals*, Physica A *388*, 1181–1186, doi: 10.1016/j.physa.2008.12.029.

Timashev, S. F. (2006), *Flicker noise spectroscopy and its application: Information hidden in chaotic signals*, Russ. J. Electrochem. *42*, 424-466, doi: 10.1134/S102319350605003X.

Timashev, S. F., Panischev, O. Yu., Polyakov, Yu. S., Demin, S. A., and Kaplan, A. Ya. (2012), *Analysis of cross-correlations in electroencephalogram signals as an approach to proactive diagnosis of schizophrenia*, Physica A *391*, 1179-1194, doi: 10.1016/j.physa.2011.09.032.

Timashev, S. F., Polyakov, Yu. S., Yulmetyev, R. M., Demin, S. A., Panischev, O. Yu., Shimojo, S., and Bhattacharya, J. (2009), *Analysis of biomedical signals by flicker-noise spectroscopy: Identification of photosensitive epilepsy using magnetoencephalograms*, Laser Phys. *19*, 836-854, doi: 10.1134/S1054660X09040434.

Timashev, S. F., Polyakov, Yu. S., Yulmetyev, R. M., Demin, S. A., Panischev, O. Yu., Shimojo, S., and Bhattacharya, J. (2010b), *Frequency and phase synchronization in neuromagnetic cortical responses to flickering-color stimuli*, Laser Physics *20*, 604-617, doi: 10.1134/S1054660X10050208.

Timashev, S.F. (2007), *Fliker-Shumovaya Spektroskopiya: Informatsiya v khaoticheskikh signalakh* (Flicker-Noise Spectroscopy: Information in Chaotic Signals), Fizmatlit, Moscow.

Timashev, S.F. and Polyakov, Y.S., (2007), *Review of flicker noise spectroscopy in electrochemistry*, Fluct. Noise Lett. *7*, R15-R47, doi: 10.1142/S0219477507003829.

Timashev, S.F. and Polyakov, Y.S., (2008), *Analysis of discrete signals with stochastic components using flicker noise spectroscopy*, Int. J. Bifurcation Chaos *18*, 2793-2797, doi: 10.1142/S0218127408022020.

Timashev, S.F. and Vstovskii, G.V. (2003), *Flicker-noise spectroscopy in analysis of chaotic time series of dynamic variables and the problem of signal-to-noise ratio*, Elektrokhimiya *39*, 149.

Timashev, S.F., Polyakov, Y.S., Misurkin, P.I., and Lakeev, S.G. (2010a), *Anomalous diffusion as a stochastic component in the dynamics of complex processes*, Phys. Rev. E *81*, 041128, doi: 10.1103/PhysRevE.81.041128.

Tributsch, H., *When the Snakes Awake: Animals and Earthquake Prediction* (MIT Press, Cambridge, MA, 1984).

Uyeda, S., Nagao, T., and Kamogawa, M. (2009), *Short-term earthquake prediction: Current status of seismo-electromagnetics*, Tectonophysics *470*, 205–213, doi: 10.1016/j.tecto.2008.07.019.

Vstovsky, G.V., Descherevsky, A.V., Lukk, A.A., Sidorin, A.Ya., Timashev, S.F. (2005), *Search for electric earthquake precursors by the method of Flicker-noise spectroscopy*, Izvestiya: Phys. Solid Earth *41*, 513-524;

Wyss, M., Aceves, R.L., Park, S.K., Geller, R.J., Jackson, D.D., Kagan, Y.Y., and Mulargia, F. (1997), *Cannot earthquakes be predicted?*, Science *278*, 487–490, doi: 10.1126/science.278.5337.487.

Wyss, M., and Booth, D.C. (1997), The IASPEI procedure for the evaluation of earthquake precursors, Geophys. J. Int. 131, 423–424, doi: 10.1111/j.1365-246X.1997.tb06587.x.

Yokoi, S., Ikeya, M., Yagi, T., and Nagai, K. (2003), *Mouse circadian rhythm before the Kobe earthquake in 1995*, Bioelectromagnetics *24*, 289–291, doi: 10.1002/bem.10108.